\documentclass[aps,pra,letterpaper,twocolumn]{revtex4-2}
\pdfoutput=1
\usepackage[utf8]{inputenc}
\usepackage[english]{babel}
\usepackage[T1]{fontenc}
\usepackage{amsmath}
\usepackage[colorlinks=true,linkcolor=blue,citecolor=magenta,urlcolor=blue]{hyperref}
\usepackage{tikz}
\usepackage{lipsum}

\begin{document}
	

\title{Minimum optical depth multiport interferometers for approximating arbitrary unitary operations and pure states}



\author{Luciano~Pereira}
\email{luciano.pereira@icfo.eu}
\affiliation{Institut de Ciencies Fotoniques, The Barcelona Institute of Science and Technology, 08860 Castelldefels, Barcelona, Spain}
\author{Alejandro~Rojas}
\affiliation{Departamento de F\'{\i}sica, Universidad de Concepci\'on, Casilla 160-C Concepci\'on, Chile}
\affiliation{Millennium Institute for Research in Optics, Universidad Concepci\'on, Concepci\'on, Chile}
\author{Gustavo~ Ca\~nas}
\affiliation{Departamento de F\'{\i}sica, Universidad del B\'{\i}o-B\'{\i}o, Collao 1202, Casilla 5C, Concepci\'on, Chile}
\affiliation{Millennium Institute for Research in Optics, Universidad Concepci\'on, Concepci\'on, Chile}
\author{Gustavo~Lima}
\affiliation{Departamento de F\'{\i}sica, Universidad de Concepci\'on, Casilla 160-C Concepci\'on, Chile}
\affiliation{Millennium Institute for Research in Optics, Universidad Concepci\'on, Concepci\'on, Chile}
\author{Aldo~Delgado}
\affiliation{Departamento de F\'{\i}sica, Universidad de Concepci\'on, Casilla 160-C Concepci\'on, Chile}
\affiliation{Millennium Institute for Research in Optics, Universidad Concepci\'on, Concepci\'on, Chile}
\author{Adán~Cabello}
\email{adan@us.es}
\affiliation{Departamento de F\'{\i}sica Aplicada II, Universidad de Sevilla, 41012 Sevilla, Spain}
\affiliation{Instituto Carlos~I de F\'{\i}sica Te\'orica y Computacional, Universidad de Sevilla, 41012 Sevilla, Spain}
	

\begin{abstract}
Reconfigurable devices which can implement arbitrary unitary operations are crucial for photonic quantum computation, optical neural networks, and boson sampling. Here, we address the problem of, using multiport interferometers, approximating with a given infidelity any pure state preparation and any unitary operation. By means of numerical calculations, we show that pure states, in any dimension $d$, can be prepared with infidelity $\le 10^{-15}$ with three layers of $d$-dimensional Fourier transforms and three layers of configurable phase shifters. We also present numerical evidence that $d+1$~layers of $d$-dimensional Fourier transforms and $d+2$~layers of configurable phase shifters can produce any unitary operation with infidelity $\le 10^{-14}$. The conclusions are achieved by numerical simulations in the range from $d = 3$ to $d=10$. These results are interesting in light of the recent availability of multicore fiber-integrated multiport interferometers.
\end{abstract}
	

\maketitle


\section{Introduction}


A programmable universal multiport array (PUMA) \cite{RZBB94,CHMKW16,LLM19} is an interferometer composed of multiple beam splitters (BSs) and phase shifters (PSs), which can be reconfigured to implement any unitary operation in a $d$-dimensional complex Hilbert space. Together with single-photon sources and detectors, PUMAs allow for preparing any quantum state, implementing any quantum logic gate and any arbitrary quantum measurement. Consequently, PUMAs are of fundamental interest for quantum computation with photons \cite{KLM01,Nielsen04,LBAJRRPOGW09,Arrazola}, optical neural networks for machine learning \cite{SHS17,SOEC19}, boson sampling \cite{BGCOSS19}, and high-dimensional entanglement \cite{MZ1997}.

\begin{figure*}[tb]
\includegraphics[width=\linewidth]{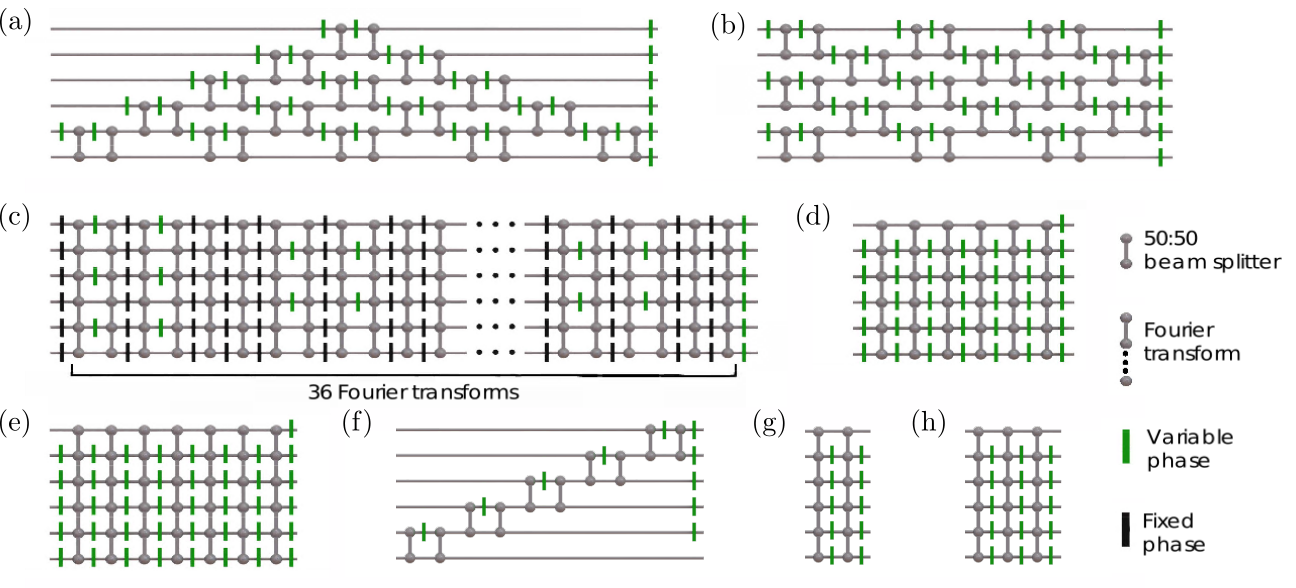}
\caption{Schemes for preparing arbitrary unitary operations and pure states in $d=6$. Every vertical segment with $d$~points represents a $d$-path Fourier transform. A Fourier in $d=2$ is a $50:50$ BS. (a)~Design of Reck {\em et al.} \cite{RZBB94}. (b)~Design of Clements {\em et al.} \cite{CHMKW16}. (c)~Design of L\'opez-Pastor, Lundeen, and Marquardt \cite{LLM19}. (d)~Design of Saygin {\em et al.} \cite{SKDSKM19}. (e)~Design proposed in this article to prepare arbitrary unitary operations. (f)~Design formed by a subset of operations from the schemes in \cite{RZBB94,CHMKW16}. (g)~Design formed with two layers of FTs. (h)~Design formed with three layers of FTs proposed in this article for pure state preparation. (a)--(h)~have optical depths $2(2d-3)$, $2d$, $6d$, $d$, $d+1$, $2(d-1)$, $2$, and $3$, respectively. 
\label{Fig1}}	
\end{figure*}

A well-known example of a PUMA is the scheme of Reck {\em et al.} \cite{RZBB94}, consisting of a regular arrangement of $d(d-1)$ $50:50$ BSs and $d^2$ tunable PSs aligned in $2(2d-3)$ layers, as illustrated in Fig.~\ref{Fig1}(a). Integrated photonics \cite{WSLT19} allowed implementing this scheme up to $d=6$ \cite{CHSM15}.

The number of layers in the interferometric array, or equivalently, the maximum number of BSs that a photon must pass through, is known as the optical depth $N$. The performance of multiport arrays is reduced by optical loss, which is directly proportional to $N$. Therefore, a smaller $N$ implies a higher quality of the implemented unitary. Clements {\em et al.} \cite{CHMKW16} noticed that the BSs and PSs can be rearranged in a configuration with $N=2d$, shown in Fig.~\ref{Fig1}(b), which has a lower depth than the scheme of Reck \emph{et al.} and therefore, is more robust to noise. This scheme has been implemented up to $d=8$ \cite{TWLE19}. 

Recent advances on multicore optical fiber fabrication technology \cite{WRWZ96,SVAMSCRO13,GL20,GL21,KMWZZ95} have made possible the realization of high-quality $d\times d$ multicore fiber multiport BSs. For $d=4$ and $d=7$, fidelities of 0.995$\pm$0.003 and 0.992$\pm$0.008, respectively, were achieved \cite{GL20} with commercially available multicore fibers. In addition, multicore fibers and related technologies were used for multidimensional entanglement generation \cite{GL21_2}, and for demonstrating the computational advantage from the quantum superposition of multiple temporal orders for higher dimensions \cite{GL21_3}. In the particular case of $d=4$, the generated multiport BS corresponds to a Hadamard matrix. In parallel, the availability of multiport BSs of $d$ ports has stimulated the search for PUMAs with small optical depth based on sequences
\begin{equation}
V = P_{N} T_d P_{N-1} T_d \cdots P_1 T_d P_{0},
\label{Sequence}
\end{equation}
where $T_d$ is the transfer matrix of the multiport beam splitter and $P_j$ corresponds to a layer of PSs introducing a different phase in each mode, that is,
\begin{equation}
P_j = \text{diag} (e^{i\phi_{j_1}}, e^{i\phi_{j_2}},e^{i\phi_{j_3}},\ldots,e^{i\phi_{j_{d}}}),
\label{PhaseShifter}
\end{equation}
with $\phi_{j_1}=0$ for $j\leq N$. Tang \emph{et al.} \cite{TTN17} numerically showed that, with a particular $T_d$ introduced in \cite{BBM94} and $N>d$, ``[any] desired
unitary matrix was generated with mean square error smaller than $-20$ dB for all tested cases.'' Zhou \emph{et al.} \cite{ZWH18} pointed out that when $T_d$ is the Fourier transform (FT) matrix in dimension $d$ (implemented by a ``canonical multiport beam splitter'' \cite{KMWZZ95}),
\begin{equation}
F_d = \frac{1}{\sqrt{d}} \sum_{j,k=1}^d e^{2 \pi i (j-1)(k-1)/d} |j\rangle\langle k |,
\end{equation}
then a sufficiently large $N$ can approximate any arbitrary unitary in dimension $d$. The question, in both cases, is which is the {\em minimum} $N$ needed to obtain universality.

Recently, it was proven that universality can be achieved with $6d + 1$ phase layers and $6d$ Fouriers \cite{LLM19}. See Fig.~\ref{Fig1}(c). However, this design has a larger depth than previous configurations $N=6d$, and then is less robust against imperfections.

Also recently, a very compact scheme was proposed by Saygin {\em et al.} \cite{SKDSKM19} consisting of a sequence of $d + 1$ layers of PSs and $d$ layers of a randomly chosen unitary transformation $T_d$ (i.e., with $N=d$). See Fig.~\ref{Fig1}(d). This scheme has two important properties: its number of PSs agrees with the number of parameters of a general unitary transformation, and it is universal and error insensitive \cite{SKDSKM19}. This last property was demonstrated using numerical simulations in $d\ge10$. In contrast to previous PUMAs \cite{RZBB94,CHMKW16,LLM19}, which can be programed following an algorithm, the scheme in \cite{SKDSKM19} requires solving a global optimization problem to derive the best settings for the PSs.


Nonuniversal multiport arrays also can find applications in particular tasks. In these cases, the multiports have lower depth, are more robust against noise \cite{TTUSTN}, and require less tunable phases. An example is the preparation of arbitrary pure states in a given dimension. For that aim it is enough to use a subset of the scheme of Reck {\em et al.}, as shown in Fig.~\ref{Fig1}(f), which has an optical depth of $2(d-1)$. The same subset of operations is also contained in the design by Clements {\em et al.} 


In this article, we address the problems of approximating with a given fidelity any unitary transformation and any pure state using devices composed of several FT and PS layers. We call a multiport array with these characteristics as pseudouniversal. Our main result is numerical evidence that the scheme with $d+1$~layers of $d$-dimensional Fourier transforms and $d+2$~layers of configurable phase shifters shown in Fig.~\ref{Fig1}(e) can produce any unitary transformation with infidelity $\le 10^{-14}$. This result improves the scheme based on $d + 1$ layers of PSs and $d$ layers of a randomly chosen unitary transformation $T_d$ \cite{SKDSKM19}, which according to our numerical experiments does not exhibit pseudouniversality in dimension $d\le 10$. In particular, this scheme generates block-diagonal unitary transformations with infidelity in the order of $10^{-7}$. Then, we show that an array with just three layers of $d$-dimensional FTs and three layers of configurable PS, as shown in Fig.~\ref{Fig1}(h), can prepare any pure state with infidelity in the order of $10^{-15}$ in any dimension from $d\le10$. A similar scheme with just two layers of FTs and PSs only achieves infidelity in between the order of $10^{-2}$ and $10^{-7}$.


\section{General considerations} 


We define the programmable pseudouniversal multiport arrays (PPUMAs) as programmable multiport arrays that allow implementing any pure state or any unitary transformation with an infidelity of at least order $O(\epsilon)$. The infidelity between two pure states is given by
\begin{align}
I(|\psi\rangle,|\phi\rangle)=1-|\langle\psi|\phi\rangle|^2.
\label{InfStates}
\end{align}
Analogously, the infidelity between two unitary transformations is given by
\begin{align}
I(U,V)=1-\frac{1}{d^2}|\mathrm{tr}(U^\dagger V)|^2.
\label{InfUnitary}
\end{align}
The definition of pseudouniversality establishes a hierarchy in PPUMAs, since PPUMAs with lower $\epsilon$ are better candidates for true universality. We focus on the lower dimensional regime ($d=3,\dots,10$), since it is still possible to carry out extensive numerical simulations covering a substantial part of the pure state and unitary transformation spaces.


\begin{figure*}[t]
\centering
\includegraphics[scale=0.34]{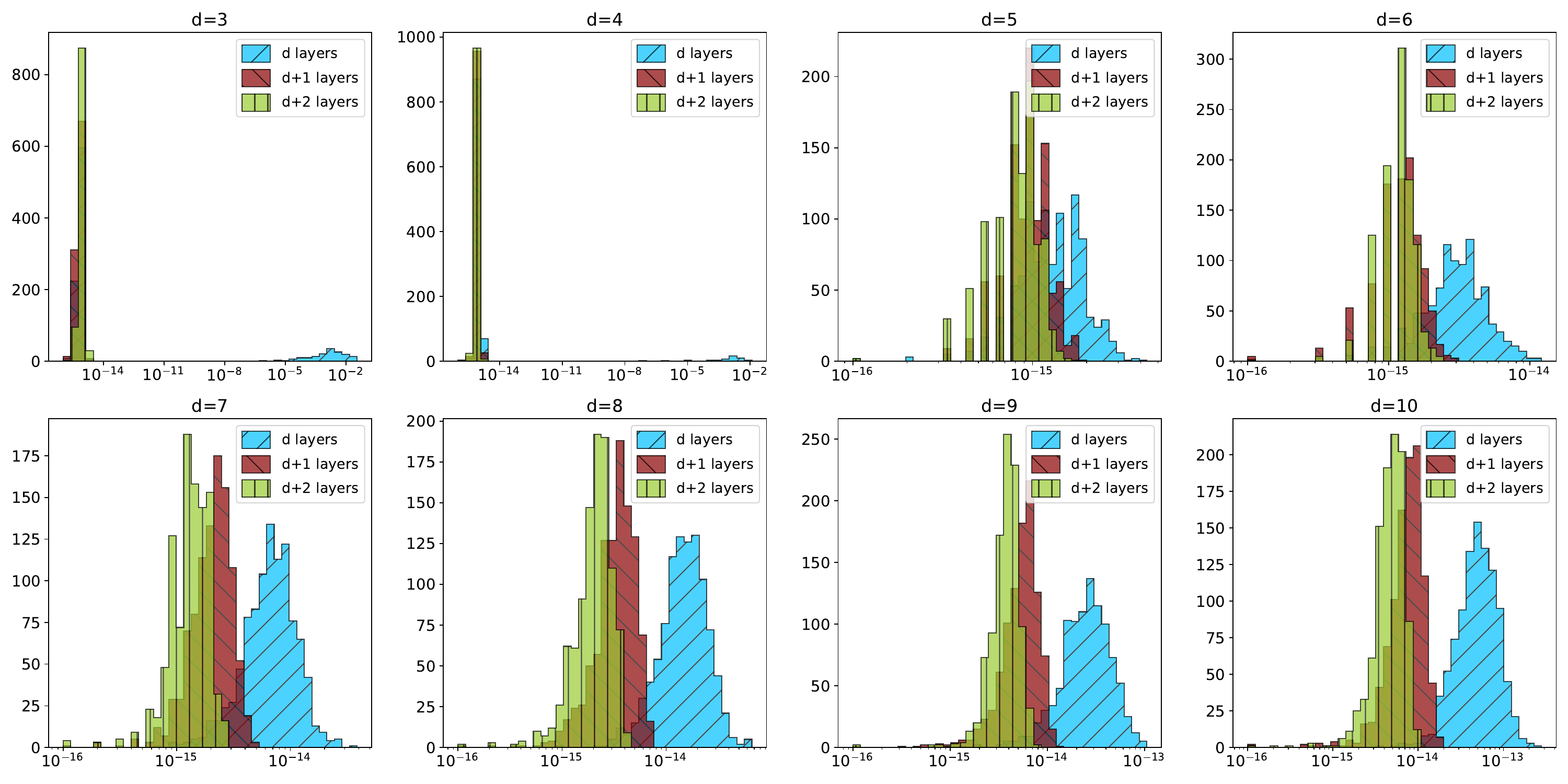}
\caption{Histograms of the logarithm of the infidelities obtained when constructing random unitary matrices from a Haar-uniform distribution using $d$ layers of FTs (and $d+1$ layers of PSs), as proposed by Saygin {\em et al.} (blue), $d+1$ layers of FTs (and $d+2$ layers of PSs), as we propose in this article (brown), and $d+2$ layers of FTs (and $d+3$ layers of PSs) (green) from dimension $d=3$ to $d=10$. For each dimension, we use $10^3$ matrices. In $d=3$ and $d=4$, around $5\%$ of the matrices cannot be constructed with high fidelity with the scheme of Saygin {\em et al.} The problem is fixed by adding an extra layer of Fouriers and PSs.}
\label{Fig2}
\end{figure*}


To study the pseudouniversality of multiport arrays, we randomly generate, according to a Haar-uniform distribution, thousands of elements from the space of pure states and from the space of unitary transformations. Afterward, we minimize the relevant infidelity to find the angles of the PSs in Eqs. (\ref{Sequence}) and (\ref{PhaseShifter}) that best approximates each element generated. 

Since the infidelities $I(|\psi\rangle,|\phi\rangle)$ and $I(U,V(\vec{\phi}))$ involve trigonometric functions, they have many local minimums, so a global optimization method has to be used. We adopt a multi-starting strategy where optimization for a given element (pure state or unitary transformation) is executed for a large set of different initial conditions to find multiple local minimal. The best local minimum is chosen as the optimal solution. The optimization problem is solved via \textsc{Julia Optim} \cite{Julia}.


\begin{figure*}[t]
\centering
\includegraphics[scale=0.34]{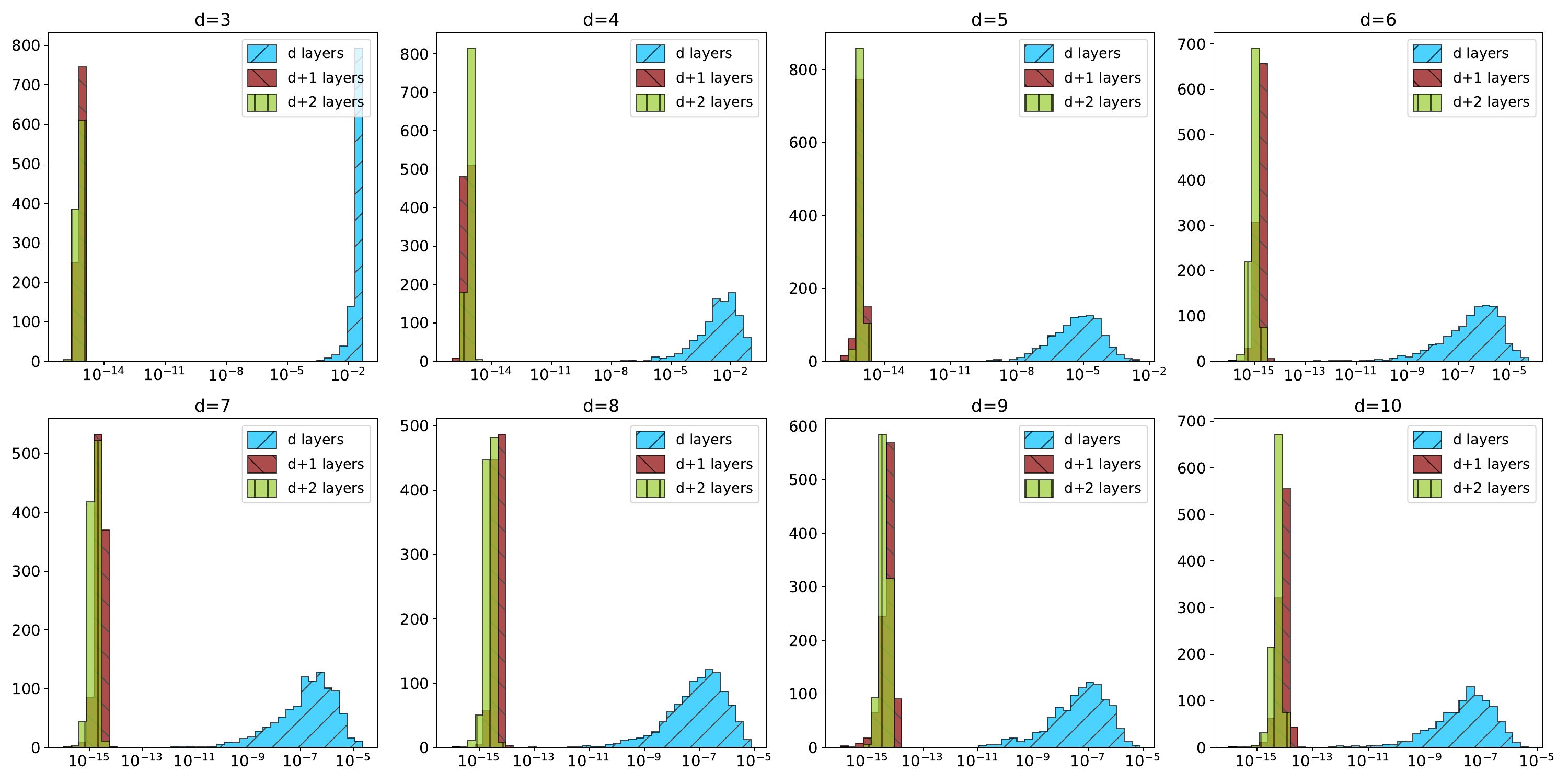}
\caption{Histograms of the logarithm of the infidelities obtained when simulating random unitary matrices in the form (\ref{form}) using $d$ layers of FTs (and $d+1$ layers of PSs), as proposed by Saygin {\em et al.}, $d+1$ layers of FTs (and $d+2$ layers of PSs), as we propose here, and $d+2$ layers of FTs (and $d+3$ layers of PSs), from dimension $d=3$ to $d=10$. Both matrices $U_{d_1}$ and $U_{d_2}$ are generated from Haar-uniform distribution. For each dimension, $10^3$ block-diagonal matrices were employed.}
\label{Fig3}
\end{figure*}


\section{Generation of arbitrary unitary transformations}


Here, we address the problem of generating arbitrary unitary transformations. For that, we compare the performance of three schemes which are candidates to be PPUMA: (i) $d + 1$ layers of PSs and $d$ layers of FTs \cite{SKDSKM19}, (ii) $d + 2$ layers of PSs and $d + 1$ layers of FTs, and (iii) $d + 3$ layers of PSs and $d + 2$ layers of FTs.


\begin{figure*}[t]
 \centering
 \includegraphics[scale=.34]{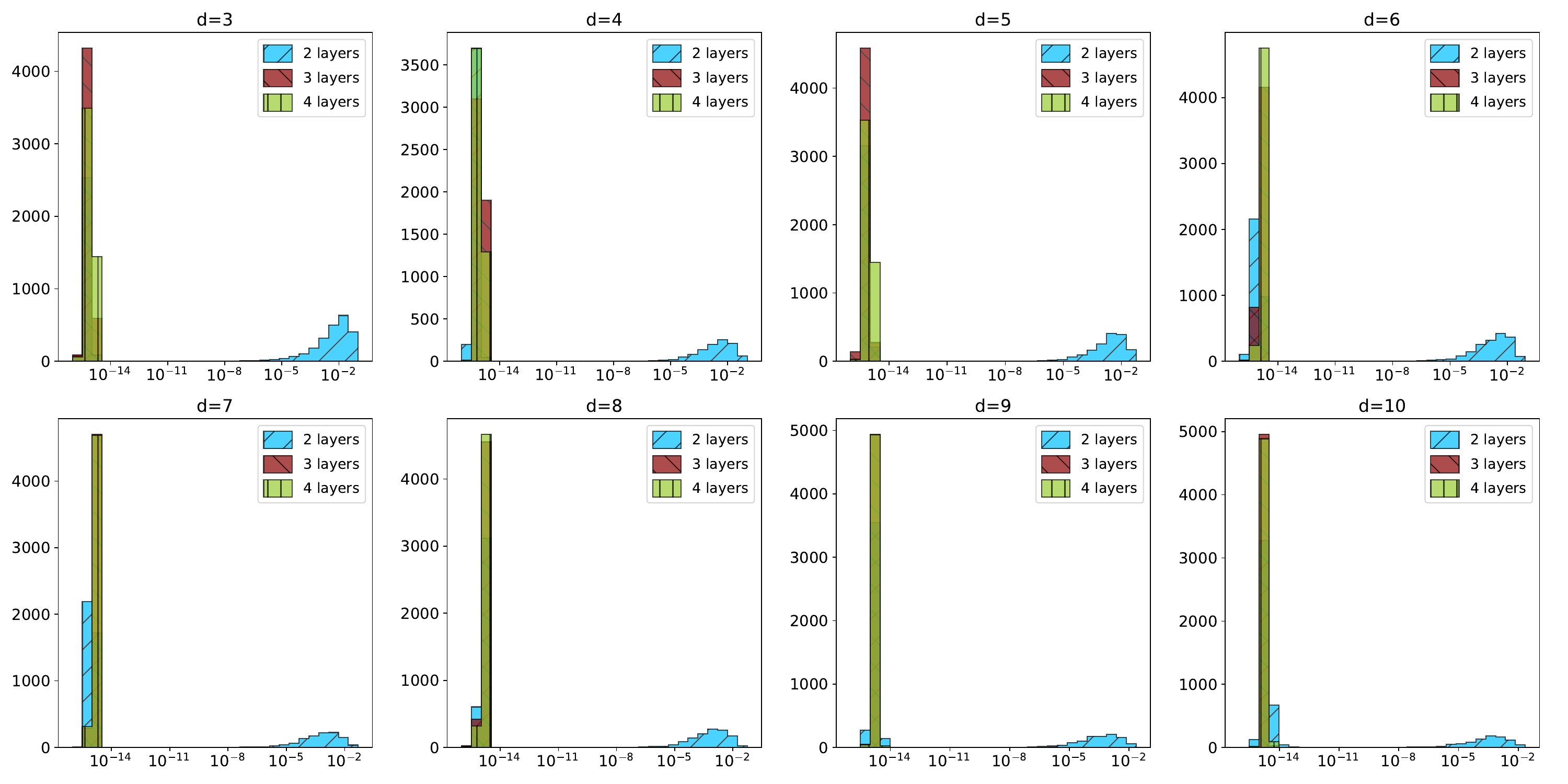}
 \caption{Histograms of the logarithm of the infidelities obtained when constructing random pure states from a Haar-uniform distribution using two, three, and four layers of FTs and PSs, from dimension $d=3$ to $d=10$. For each dimension, we used $5\times10^3$ states. The case of two layers exhibits a set of states generated with infidelity in the order of $O(10^{-5})$, in the best case, in all dimensions. This set disappears when using a configuration with three layers, which consistently leads to infidelities in the order of $O(10^{-15})$.}
 \label{Fig4}
\end{figure*}


We explore all dimensions from $d=3$ to $d=10$. For each arrangement in each dimension, we try $10^3$ unitary matrices from a Haar-uniform distribution. In all dimensions, $30$ randomly generated initial conditions are employed. The results of the optimization are shown in Fig.~\ref{Fig2}, where histograms for the best-achieved infidelity (in logarithmic scale) are exhibited. We can see that, in $d=3$ and $4$, the use of $d+1$ layers of PSs and $d$ layers of FTs leads to approximately a $5\%$ of matrices that cannot be constructed with high fidelity. These are constructed with an infidelity of order $O(10^{-2})$. This can be due to two reasons: either the configuration is a PPUMA with error $O(10^{-2})$, or the optimization algorithm was unable to find the global minimum. To rule out this last possibility, this $5\%$ of matrices is optimized again with a larger set of initial conditions. However, no significant reduction in infidelity is achieved. In addition, numerical simulations are also implemented using \textsc{Matlab GlobalSearch} and \textsc{Python Basin-Hopping}. These maintain the main features exhibited in Fig.~\ref{Fig2} but an overall increase in the infidelity for the three tests. This indicates the existence of unitary transformations that cannot be accurately implemented by means of $d+1$ layers of PSs and $d$ layers of FTs, and moreover, that this configuration is a PPUMA $O(10^{-2})$ and not completely universal for $d=3$ and $d=4$. Interestingly, it can be seen that for $d+1$ and $d+2$ layers of FTs, the implementation of the same matrices is typically done with better infidelity than the one achieved by means of $d$ layers, being the configurations with $d+1$ and $d+2$ FTs at least PPUMA $O(10^{-14})$. However, from Fig.~\ref{Fig2} we can see that the fidelities delivered by means of $d+1$ layers of PSs and $d$ layers of FTs in $d=5,6,\ldots,10$ are worse, at most, by one order of magnitude with respect to the other schemes. However, this difference can be considered marginal as the dimension increases. This suggests that all schemes are PPUMA $O(10^{-13})$ in high dimensions.

Randomly chosen unitary matrices according to the Haar distribution uniformly represent the space of unitaries. However, there are matrices that will never appear in the sample since they span a subspace of null measure. To check the universality of the scheme, we look for this type of subset whose elements cannot be constructed with high fidelity. Specifically, we consider matrices of the form
\begin{equation}
\label{form}
U = U_{d_1} \oplus U_{d_2} = \begin{pmatrix}
U_{d_1} & 0 \\ 0 & U_{d_2}
\end{pmatrix},
\end{equation}
where $U_{d_j}$ is a unitary matrix of size $d_j$. We choose $d_j = d/2$ for $d$ even and $d_1= (d-1)/2$ and $d_2 = (d + 1) / 2$ for $d$ odd. Figure~\ref{Fig3} presents histograms of the infidelity (in logarithmic scale) that result from optimizing $10^3$ randomly generated block-diagonal unitary transformations of the form (\ref{form}) from a Haar-uniform distribution with $d=3,\dots,10$. The optimization is done as explained before. In the case of $d$ layers of FTs, the histograms are spread along two to four orders of magnitude of the infidelity and achieve values that are at least two orders of magnitude worse than the case of $d+1$ layers of FTs. Particularly, for $d = 3,4$ the infidelities are order $O(10^{-2})$, while for $d = 5$ to $d = 10$ they are at most $O(10^{-5})$. These last results do not coincide with those obtained on unitary matrices in Haar distribution, where all were reconstructed with infidelities of order $O(10^{-13})$. However, infidelities obtained with $d+1$ FTs are concentrated in a narrow interval around infidelities of order $O(10^{-14})$. Furthermore, the histograms obtained with $d+1$~layers in Figs.~\ref{Fig2} and \ref{Fig3} are very similar.

All of this indicates that the configuration based on $d+ 1$ layers of PSs and $d$ layers of FTs is a PPUMA $O(10^{-5})$, and that supplementing it with an additional layer allows us to implement a PPUMA $O(10^{-14})$. Clearly, the addition of an extra layer of FT and PS leads to a large increase in accuracy of the implemented unitary transformation for block-diagonal unitary transformations. This leads to the question of whether adding more layers could lead to even better accuracy. To examine this, we carry out a third test adding an extra layer of FTs and PSs. Simulations with this later configuration do not exhibit a significant increase in accuracy with respect to the case of $d+1$ layers of FTs. Therefore, we have that the scheme with $d+1$ layers of FT and $d+2$ layers of PS is the PPUMA with lower depth.

Since the scheme with $d+1$ layer of FTs has $d-1$ more phases than is needed to characterize a unitary transformation in $U(d)$, we try three strategies to reduce the number of phases: (i)~deleting a randomly chosen phase, (ii)~fixing $d-1$~phases at the inner layers, and (iii)~fixing $d-1$~phases of the last layer. In all cases, simulations show that each intervention conveys a decrease in pseudouniversality.


\section{Generation of arbitrary pure states}


Pure states in a $d$-dimensional Hilbert space are defined by $2d-2$ independent real parameters. This indicates that a configuration with lower numbers of layers than a PPUMA for unitary matrices could be sufficient to generate any pure state in any dimension, being the configuration with two layers of FTs, which has exactly $2d-2$ tunable phases, the smallest possible candidate, as is shown in Fig.~\ref{Fig1}(g). To test this hypothesis, we consider three different candidates for PPUMA, with two, three, and four layers of FTs and PSs. A given target state $|\phi\rangle$ is generated by applying the unitary transformation $U$ onto a fixed state $|1\rangle$, where $U$ corresponds to the action of the different layers of FTs and PSs. The infidelity $I(U|1\rangle,|\phi\rangle)$ is minimized with respect to the phases introduced by the layers of PSs. For each dimension, a set of $5\times10^3$ pure states Haar-uniform distributed was generated and each of them was reconstructed $20$ times to avoid local minimums. 

The result of the optimization procedure for two, three, and four layers is summarized in Fig.~\ref{Fig4}, which corresponds to an infidelity histogram. As it is clear from this figure, the configuration with two layers leads to two well-defined populations: a set of states that, in the best case, are generated with an infidelity in the order $O(10^{-5})$ and set of states generated with very low infidelities in the order of $O(10^{-15})$. These two sets are present across all inspected dimensions. The size of the set of high infidelity states slowly decreases with the dimension. To rule out the possibility of failure of the optimization algorithm on the states of the later set, a second optimization attempt was carried out. This time, different initial guesses were employed, and two other optimization algorithms, \textsc{Matlab GlobalSearch}, and \textsc{Python Basin-Hopping}, were used. In spite of this, our main findings remain unchanged. This strongly suggests the existence of pure states that cannot be accurately generated with a configuration based on two layers of FTs and two layers of PSs. This result can be greatly improved, as Fig.~\ref{Fig4} shows, by increasing the number of layers from two to three. In this case, all states were generated with an infidelity in the order of $O(10^{-15})$, which suggests that three layers form a PPUMA $O(10^{-15})$ for the generation of states. This result raises the question of whether a further increase in the number of layers can lead to a reduction in infidelity. To study this, we performed numerical simulations with four layers of FTs and PSs. These, however, also lead to infidelities in the order of $O(10^{-15})$ for all simulated states. This shows that the PPUMA based on three layers could provide the configuration with the smallest optical depth to generate pure states, and consequently, with the highest robustness to errors. 

The optical depth of our proposal compares favorably with other schemes. In particular, the design by Reck {\it et al.} generates any pure state employing a Mach-Zehnder interferometer between neighboring paths and a final layer of PSs, which corresponds to the first diagonal in Fig.~\ref{Fig1}(f). The same array of BSs and PSs generates any arbitrary state in the Clemens {\it et al.} configuration, which corresponds to the main anti-diagonal in Fig.~\ref{Fig1}(b). In both designs, a pure state is given as a lineal combination of the basis states where the probability amplitudes are given by complex phases multiplied by hyperspherical coordinates. This leads to an optical depth of $2(d-1)$, which is greater than the optical depth of 3 achieved by our proposal.


\section{Conclusions}


For photonic universal quantum computation and many applications, it is of fundamental importance to identify the PUMA with the smallest optical depth. In this article, we advance the solution of this problem by considering multiport arrays based on Fourier transform through several Monte Carlo numerical simulations for $d=3,\dots,10$. We show that an array composed of $d+1$ layers of FTs and $d+2$ layers of PSs is a PPUMA that generates arbitrary unitary transformations with an infidelity of at least order $O(10^{-14})$. This configuration outperforms the infidelity achieved using $d$ layers of FTs and $d+1$ layers of PSs, of $O(10^{-2})$ for random unitary matrices in $d=3,4$ (see Fig.~\ref{Fig2}), and in between $O(10^{-2})$ and $O(10^{-7})$ for block diagonal unitary transformations in all studied dimensions (see Fig.~\ref{Fig3}), in several order of magnitudes. Furthermore, we show that increasing the number of FT layers to $d+2$ does not result in any significant improvement in infidelity (see Figs.~\ref{Fig2} and \ref{Fig3}). The scheme in \cite{SKDSKM19} with $d$ layers has phases that match the number of parameters of a unitary operation, so it is the natural candidate for minimal PUMA. However, simulations suggest that some redundancies exist that reduce the number of independent parameters, so not all unitaries can be approximated with high fidelity. This scheme would be good enough for some purposes, but it is not a high-fidelity PPUMA. All of these point out that the best candidate for the PPUMA with minimal optical depth may be the one with $d + 1$ layers of FTs and $d + 2$ layers of PSs, as shown in Fig.~\ref{Fig1}(e). Introducing an extra layer of phases increases the number of independent parameters, obtaining much better accuracy for general purposes. Nevertheless, the problem of the PUMA with the minimum depth remains open.

We also studied the problem of generating pure states through multiport arrays based on layers of FTs and PSs. Since pure states have $2d-2$ independent real parameters, a first guess is a configuration based on two layers of FTs and two layers of PSs. However, we show that this configuration generates a set of states, in the best case, with an infidelity in the order of $O(10^{-5})$. Another set of states is generated with very low infidelities in the order of $O(10^{-15})$. These two sets are present across all inspected dimensions. Increasing the number of layers to 3 we obtain a configuration that consistently generates all states with an infidelity in the order of $O(10^{-15})$ (see Fig.\thinspace\ref{Fig4}). As in the case of unitary transformations, a further increase to four layers does not lead to a decrease in the infidelity. This shows that the PPUMA based on three layers could provide the configuration with the smallest optical depth and, consequently, with the highest robustness to errors. 

As long as no universal PUMA from an analytical argument is found, it is important to continuously numerically study the problem to find new sets of unitary matrices that are implemented with low infidelity with the tested configurations, and thus fine-tune the best candidate to be universal.
A direct avenue of our work is the approximation of unitaries and pure states in a noisy context. If the multiport beam splitters and the phase shifters are noisy, they do not implement the sequence perfectly of Eq.~\eqref{Sequence}. This clearly would reduce the fidelity of the implementation of the unitaries, requiring extending the study to the quantum process formalism \cite{28}. Here, we would employ quantum process tomography \cite{29} to obtain noisy representations of the multiport beam splitters and phase shifters. Then, we optimize the fidelity between the target unitary matrix and the noisy operation to search for the best set of phases to approximate the target unitary matrix. The second avenue is to extend the study to higher dimensions. Currently, multiport BSs have increased to $19$ modes \cite{30}, and through concatenations of them, interferometers of even larger dimensions could be designed. A similar numerical study can be done on these higher-dimensional devices. Nevertheless, given the size of the operations and the number of phases, the optimization of the fidelity is far harder. Thereby, it would require the use of or the development of new global optimization techniques to efficiently find the optimum.


\section*{Data Availability}


The data and codes to reproduce Figs.~\ref{Fig2}, \ref{Fig3}, and \ref{Fig4} are publicly available in Github \cite{github_data}.


\begin{acknowledgments}
The authors thank Marco T\'ulio Quintino and Giuseppe Vallone for conversations. LP is supported by the Government of Spain (Severo Ochoa CEX2019-000910-S, FUNQIP, Quantum in Spain, and European Union NextGenerationEU PRTR-C17.I1), Fundació Cellex, Fundació Mir-Puig, and Generalitat de Catalunya (CERCA program). AD, GC, and GL are supported by FONDECYT Grants No.~1180558, No.~1230796, and No.~1160400, respectively. AC is supported by 
the EU-funded project \href{10.3030/101070558}{FoQaCiA} and the \href{10.13039/501100011033}{MCINN/AEI} (Project No.~PID2020-113738GB-I00).
\end{acknowledgments}



\begin{thebibliography}{99}


\bibitem{RZBB94}
M. Reck, A. Zeilinger, H. J. Bernstein, and P. Bertani, Experimental realization of any discrete unitary operator, \href{https://doi.org/10.1103/PhysRevLett.73.58}{Phys. Rev. Lett. {\bf 73}, 58 (1994).}

\bibitem{CHMKW16}
W. R. Clements, P. C. Humphreys, B. J. Metcalf, W. S. Kolthammer, and I. A. Walmsley, Optimal design for universal multiport interferometers, \href{https://doi.org/10.1364/OPTICA.3.001460}{Optica {\bf 3}, 1460 (2016).}

\bibitem{LLM19}
V. J. L\'opez-Pastor, J. S. Lundeen, and F. Marquardt, Arbitrary optical wave evolution with Fourier transforms and phase masks, \href{https://doi.org/10.1364/OE.432787}{Opt. Express {\bf 29}, 38441 (2021).}

\bibitem{KLM01}
E. Knill, R. Laflamme, and G. J. Milburn, A scheme for efficient quantum computation with linear optics, \href{https://doi.org/10.1038/35051009}{{\it Nature} {\bf 409}, 46 (2001).}

\bibitem{Nielsen04}
M. A. Nielsen, Optical Quantum Computation Using Cluster States, \href{https://doi.org/10.1103/PhysRevLett.93.040503}{Phys. Rev. Lett. {\bf 93}, 040503 (2004).}

\bibitem{LBAJRRPOGW09}
B. P. Lanyon, M. Barbieri, M. P. Almeida, T. Jennewein, T. C. Ralph, K. J. Resch, G. J. Pryde, J. L. O'Brien, A. Gilchrist, and A. G. White, Simplifying quantum logic using higher-dimensional Hilbert spaces, \href{https://doi.org/10.1038/nphys1150}{Nat. Phys. {\bf 5}, 134 (2009). }

\bibitem{Arrazola} J. M. Arrazola {\it et al.}, Quantum circuits with many photons on a programmable nanophotonic chip, \href{https://doi.org/10.1038/s41586-021-03202-1}{Nature {\bf 591}, 54 (2021).}

\bibitem{SHS17}
Y. Shen, N. C. Harris, S. Skirlo, M. Prabhu, T. Baehr-Jones, M. Hochberg, X. Sun, S. Zhao, H. Larochelle, D. Englund, and M. Solja\v{c}i\'{c}, Deep learning with coherent nanophotonic circuits, \href{https://doi.org/10.1038/nphoton.2017.93}{Nat. Photonics {\bf 11}, 441 (2017).}

\bibitem{SOEC19}
G. R. Steinbrecher, J. P. Olson, D. Englund, and J. Carolan, Quantum optical neural networks, \href{https://doi.org/10.1038/s41534-019-0174-7}{npj Quant. Inf. {\bf 5}, 60 (2019).}

\bibitem{BGCOSS19}
K. Mattle, M. Michler, H. Weinfurter, A. Zeilinger, and M. \.Zukowski, Nonclassical statistics at multiport beam-splitters, Appl. Phys. B {\bf 60}, S111, 1995.

\bibitem{MZ1997} 
M. \.Zukowski, A. Zeilinger, and M. A. Horne, Realizable higher-dimensional two-particle entanglements via multiport beam splitters,
\href{https://doi.org/10.1103/PhysRevA.55.2564}{Phys. Rev. A {\bf 55}, 2564 (1997).}

\bibitem{WSLT19}
J. Wang, F. Sciarrino, A. Laing, and M. G. Thompson, Integrated photonic quantum technologies, \href{https://doi.org/10.1038/s41566-019-0532-1}{Nat. Photonics {\bf 14}, 273 (2020).}

\bibitem{CHSM15}
J. Carolan, C. Harrold, C. Sparrow, E. Mart\'{\i}n-L\'opez, N. J. Russell, J. W. Silverstone, P. J. Shadbolt, N. Matsuda, M. Oguma5, M. Itoh, G. D. Marshall, M. G. Thompson, J. C. F. Matthews, T. Hashimoto, J. L. O'Brien, and A. Laing, Universal linear optics, \href{https://science.sciencemag.org/content/349/6249/711}{Science {\bf 349}, 711 (2015).}

\bibitem{TWLE19}
C. Taballione, T. A. W. Wolterink, J. Lugani, A. Eckstein, B. A. Bell, R. Grootjans, I. Visscher, D. Geskus, C. G. H. Roeloffzen, J. J. Renema, I. A. Walmsley, P. W. H. Pinkse, and K.-J. Boller, 8$\times$8 reconfigurable quantum photonic processor based on silicon nitride waveguides, \href{https://doi.org/10.1364/OE.27.026842}{Opt. Express {\bf 27}, 26842 (2019).}

\bibitem{WRWZ96}
G. Weihs, M. Reck, H. Weinfurter, and A. Zeilinger, All-fiber three-path Mach–Zehnder interferometer, \href{https://doi.org/10.1364/OL.21.000302}{Opt. Lett. {\bf 21}, 302 (1996).}

\bibitem{SVAMSCRO13}
N. Spagnolo, C. Vitelli, L. Aparo, P. Mataloni, F. Sciarrino, A. Crespi, R. Ramponi, and R. Osellame, Three-photon bosonic coalescence in an integrated tritter, \href{https://doi.org/10.1038/ncomms2616}{Nat. Commun. {\bf 4}, 1606 (2013).}

\bibitem{GL20} 
J. Cari\~ne, G. Ca\~nas, P. Skrzypczyk, I. \v Supi\'c, N. Guerrero, T. Garcia, L. Pereira, M. A. S.-Prosser, G. B. Xavier, A. Delgado, S. P. Walborn, D. Cavalcanti, and G. Lima, Multi-core fiber integrated multiport beam splitters for quantum information processing, \href{https://doi.org/10.1364/OPTICA.388912}{Optica {\bf 7}, 542 (2020).}

\bibitem{GL21} 
M. Farkas, N. Guerrero, J. Cari\~ne, G. Ca\~nas, and G. Lima, Self-Testing Mutually Unbiased Bases in Higher Dimensions with Space-Division Multiplexing Optical Fiber Technology, \href{https://link.aps.org/doi/10.1103/PhysRevApplied.15.014028}{Phys. Rev. Applied {\bf 15}, 014028 (2021).}

\bibitem{KMWZZ95}
J. Zhou, J. Wu, and Q. Hu, Tunable arbitrary unitary transformer based on multiple sections of multicore fibers with phase control, \href{https://doi.org/10.1364/OE.26.003020}{Opt. Express {\bf 26}, 3020 (2018).}

\bibitem{GL21_2}
 E. S. G\'omez, S. G\'omez, I. Machuca, A. Cabello, S. P\'adua, S. P. Walborn, and G. Lima, Multidimensional Entanglement Generation with Multicore Optical Fibers, 
\href{https://link.aps.org/doi/10.1103/PhysRevApplied.15.034024}{Phys. Rev. Applied {\bf 15}, 034024 (2021).}

\bibitem{GL21_3}
M. M. Taddei, J. Cari\~ne, D. Mart\'{\i}nez, T. Garc\'{\i}a, N. Guerrero, A. A. Abbott, M. Ara\'ujo, C. Branciard, E. S. G\'omez, S. P. Walborn, L. Aolita, and G. Lima, Computational Advantage from the Quantum Superposition of Multiple Temporal Orders of Photonic Gates, 
\href{https://link.aps.org/doi/10.1103/PRXQuantum.2.010320}{PRX Quantum {\bf 2}, 010320 (2021).}
 
\bibitem{TTN17}
R. Tang, T. Tanemura, and Y. Nakano, Integrated Reconfigurable Unitary Optical Mode Converter Using MMI Couplers, \href{https://doi.org/10.1109/LPT.2017.2700619}{IEEE Photon. Tech. Lett. 29, {\bf 2700619} (2017).}

\bibitem{BBM94}
M. Bachmann, P. A. Besse, and H. Melchior, General self-imaging properties in N$\times$N multimode interference couplers including phase relations, \href{https://doi.org/10.1364/AO.33.003905}{Appl. Opt. 33, {\bf 3905} (1994).}

\bibitem{ZWH18}
D. J. Brod, E. F. Galv\~{a}o, A. Crespi, R. Osellame, N. Spagnolo, and F. Sciarrino, Photonic implementation of boson sampling: a review, \href{https://doi.org/10.1117/1.AP.1.3.034001}{Adv. Phot. {\bf 1}, 034001 (2019).}

\bibitem{SKDSKM19}
M. Y. Saygin, I. V. Kondratyev, I. V. Dyakonov, S. A. Mironov, S. S. Straupe, and S. P. Kulik, Robust Architecture for Programmable Universal Unitaries, \href{https://doi.org/10.1103/PhysRevLett.124.010501}{Phys. Rev. Lett. {\bf 124}, 010501 (2020).}

\bibitem{TTUSTN} R. Tanomura, R. Tang, T. Umezaki, G. Soma, T. Tanemura, and Y. Nakano, Scalable and Robust Photonic Integrated Unitary Converter Based on Multiplane Light Conversion, \href{http://dx.doi.org/10.1103/PhysRevApplied.17.024071}{Phys. Rev. Appl. {\bf 17}, 024071 (2022).}

\bibitem{Julia} \href{https://julianlsolvers.github.io/Optim.jl/stable/}{https://julianlsolvers.github.io/Optim.jl/stable/}

\bibitem{28},
S. Milz, F. A. Pollock, and K. Modi, 
An introduction to operational quantum dynamics,
\href{http://dx.doi.org/10.1142/S1230161217400169}{Open Syst. Inf. Dyn. {\bf 24}, 1740016 (2017).}

\bibitem{29},
I. L. Chuang and M. A. Nielsen, Michael,
Prescription for experimental determination of the dynamics of a quantum black box,
\href{http://dx.doi.org/10.1080/09500349708231894}{J. Mod. Opt. {\bf 44}, 2455 (1997).}

\bibitem{30}
E. A. Ortega, K. Dovzhik, J. Fuenzalida, S. Wengerowsky, J. C. Alvarado-Zacarias, R. F. Shiozaki, R. Amezcua-Correa, M. Bohmann, and R. Ursin,
Experimental Space-Division Multiplexed Polarization-Entanglement Distribution through 12 Paths of a Multicore Fiber,
\href{http://dx.doi.org/10.1103/PRXQuantum.2.040356}{PRX Quantum {\bf 2}, 040356 (2021).}

\bibitem{github_data} \href{https://github.com/akiiop/PUMA}{https://github.com/akiiop/PUMA}

	
\end{thebibliography}
\end{document}